# An Efficient, Distributable, Risk Neutral Framework for CVA Calculation

Dongsheng Lu and Frank Juan

September 2010

## Abstract

*The importance of counterparty credit risk to the derivative contracts was demonstrated consistently throughout the financial crisis of 2008. Accurate valuation of Credit value adjustment (CVA) is essential to reflect the economic values of these risks. In the present article, we reviewed several different approaches for calculating CVA, and compared the advantage and disadvantage for each method. We also introduced an more efficient and scalable computational framework for this calculation.*

**1. Introduction**

The importance of counterparty credit risk to the derivative contracts was demonstrated consistently throughout the financial crisis of 2008. Credit value adjustment (CVA), which is the accounting treatment to reflect the fair value of counterparty credit risk, frequently caused large trading profit and loss (P&L) swing for some derivative houses during the crisis. While it may be tempting to consider CVA as loan loss reserve equivalent for derivatives, there is a fundamental difference; namely, CVA is intended to reflect the economic value for carrying counterparty credit risk based on the market at the particular point in time, the loan loss reserve, on the other hand, is simply the presumed objective estimate of expected credit loss. In other words, in addition to the expected loss CVA includes the current market price of risk or market's sentiment to risk. As the market for credit derivatives matures over the past decade, more and more tools are now available to the traders for managing the counter party credit risk associated with derivatives trading. Managing counterparty credit exposure no longer depends purely on setting credit risk limits according to potential future exposure. Rather, credit risk can now be hedged with various credit derivatives readily available such as CDS[1]. Managing counterparty credit risk in fact has become an integrated part of many derivative trading desks' day to day activities[2]. The cost for managing this credit risk (CVA) therefore is routinely included in the quoted prices for derivatives provided by dealers nowadays. Utilizing the information from CDS market is therefore essential for reflecting the economic value of credit cost in any CVA calculation. The accuracy and sophistication of CVA calculation is also getting ever more demanding. For Example, various detailed features of CSA (Credit Support Annex), such as mutual put breaks and automatic trigger event will need to be addressed in a CVA calculation. The economic values of these credit mitigants have to be properly accounted for so that there is proper incentive to utilize these risk mitigation techniques and the business can compete fairly in a very competitive market.

---

[1] One issue is how to hedge one's own credit risk or the corresponding economic value, DVA. The authors believe that this should be accounted for naturally in the funding cost of the trading activity in a normal market environment. There is currently however an asymmetry between CDS spread and funding spread due to the fact certain financial institutions have no need for funding from market in an environment the central bank is running a quantitative easing policy.

[2] In many financial institutions, a dedicated CVA desk was established to warehouse all [counterparty] credit risk centrally so that this risk can be managed more efficiently.

There have been several studies over the years on issues related to the calculation for counterparty credit risk exposure and CVA. For example, Canabarro and Duffie [CD03] provided an excellent introduction on measuring and valuing counterparty credit risk. "Counterparty Credit Risk Modeling" [PYK05] edited by Pykhtin provided more detailed reviews on various facets in the CVA calculations. Pykhkin and Zhu [PZ07] also gave an excellent overview on a comprehensive framework for pricing CVA.

There are two general frameworks for CVA calculation, forward simulation and backward inductions. The framework discussed in details by Pykhkin and Zhu (PZ) is what we will call forward simulation framework, where computations are done along all simulated paths going forward in time. In this framework, a large number of Monte Carlo scenarios are generated for market risk factors that impact the valuation of the derivative portfolio at discrete time steps from current time to the time when all instruments within the portfolio expire. Each individual position within the portfolio is valued at every time step along each simulated path, from which the exposure profiles of portfolio are calculated at every time step. Netting agreement, collateral thresholds and other credit mitigants are applied during the calculation of exposure profile. Assuming defaults are independent of market risk factors, the CVA can be calculated as:

$$CVA = PV((1-R_c)\sum_i E^+(t_i)dP_{d,c}(t_i) - (1-R_o)\sum_i E^-(t_i)dP_{d,o}(t_i)) \qquad (1)$$

Where $PV$ indicates present value operation, $E^+(t_i)$ and $E^-(t_i)$ are the expected positive and negative exposures respectively, $dP$ is the discrete default probability, and $R$ is the recovery rate. We use the subscript $o$, and $c$ to differentiate between the counterparty and one's own credit risk. One tricky issue within this approach is the consistency between the risk factors used for derivative valuation and the risk factors used for CVA calculation. We will discuss further about this point at later sections.

The second approach is to calculate CVA through a backward pricing framework. Within this framework, valuations of all instruments are performed twice, once in a default free setting and once in the presence of default risk. In this framework, CVA is simply the difference between these two valuations:

$$CVA = V_d - V_{nd} \qquad (2)$$

During both the default free and defaultable settings, all instruments will be valued using the same risk factors under the same pricing measure. The pricing setting could be trees, finite difference or Monte Carlo simulations.

The remaining of this article is organized as follows: first, we will discuss a little more in details about the forward simulation and backward pricing framework. Pros and cons will be considered as well as implementation details. Credit simulations with correlations to market risk factors are discussed within the framework of structural model. Different approaches to obtain a risk neutralized transition matrix and techniques to the migration in trees/lattice/Monte Carlo simulations are discussed. Afterwards, we will present an efficient CVA calculation methodology, which can be related to both frameworks. This methodology is scalable, distributable and easy to implement. Credit migration, rating based threshold and ATEs are also considered within credit simulations. We will address the sensitivity calculations for CVA, including market risk sensitivity and CDS sensitivity as well as wrong way exposure. Finally we will discuss the impact of exercise boundary decision in the presence of default and some practical issues.

## 2 Forward Simulation Framework

Forward simulation framework involves the valuations of all instruments at every point of time and every scenario. The loss profiles L(t) are obtained at every time point t, which gives the distribution of portfolio valuations at different market risk scenarios. Collateral thresholds can be applied with netted portfolio value. The maximum exposure at a given time point and scenario would be min[NettingValue, Collateral Threshold].

The computation requirement involved in forward simulation framework is very intensive: N×M calculations for all K instruments. While this might still be manageable for linear instruments with the powerful computing machines routinely employed by the top financial institutions, the brute force approach quickly becomes not economical for more complicated derivatives and exotic instruments. In order to reduce the required computation time, pricing accuracy has to be compromised by making approximations in pricing function.

To implement the forward simulation, the financial instruments will need to be aged as they are simulated through time, and the pricing functions have to be adapted to handle such aging. For example, fixings and resets of swaps and exercises of options will need to be tracked at every time step. While numeric tricks like Brownian bridge can be used to handle past fixings by interpolating rates between time points, complication arises with Bermudan exercise and/or trigger events due to their path-dependent characteristic. At each time step during the simulation a decision has to be made regarding whether the deal has been exercised or triggered previously. Additional approximations based on algorithms like conditional valuation are therefore necessary to solve this issue.

PZ's framework does not perform credit simulation, therefore it would have difficulty handling rating based collateral thresholds. For example, assuming the following CSA agreements with the following collateral thresholds against a counterparty:

  AA Threshold:      50M

  A  Threshold:      10M

  BBB Threshold:     0 M

This means that if the counterparty rating drops to A from AA, the collateral threshold will decrease from 50M to 10M, or the maximum exposure to the counterparty will be 10M; if the rating drops to BBB or below, the threshold will decrease to 0, meaning counterparty will have to post sufficient collateral to cover the current exposure (any positive MTM values). Obviously the rating based thresholds are valuable credit mitigants, and its economic value should be reflected in CVA calculations. In a forward simulation framework, this would imply we know the rating of counterparty at each scenario or the distribution of rating across scenarios, which would require the simulation of credit migrations. We defer the discussion on extension of credit simulations in this framework to a later section.

**3 Backward Pricing Framework**

Backward pricing based on default and no-default valuations provides an efficient alternative to the forward simulation framework. In this section we discuss in details how it can be implemented. In general, the portfolio of instruments are priced within the same backward pricing mechanism, regardless it is a tree, a lattice or a Monte Carlo simulation. Knowing the value of each instrument at every node/path, the netting across the portfolio can be performed easily. Defaults and migration of credit can be naturally accounted for by stepping backwards in time through discounting. We will explain the implementation in details below.

In the first step, we will generate a generic time grid for pricing all instruments. The last time grid point is selected based on the longest maturity of these instruments. The time grid is set up to includes as many cashflow, exercise/trigger schedules as possible. The grids are distributed with more time points in the front and less points in the back. Given that Cashflow at 10 years 1 week is barely distinguishable from cashflow at 10 years, fewer time grid points are selected as time goes further into the future. This should not pose any problem with the much higher tolerance in accuracy for CVA.

Based on the generic time grid established, a generate risk neutral tree or scenarios is produced as in regular derivative pricing. Either short rate tree or Libor Market model should serve the present purpose of CVA calculation for vanilla interest rate derivatives. However, for structure desk with exotic instruments priced with many different pricing models, a consistent model should be selected for the CVA calculations. For interest rate derivatives, we found that a globally calibrated cross currency LMM seems to be a good choice as most fixed income derivatives can be priced reasonably well with this model.

Before getting into details about instrument pricing, it is worthwhile to discuss the default and rating based threshold mechanism. There are two different approaches to handle the credit setting: one is to use the specific default/credit information to identify the exposure at default from scenarios, which we will discuss later. The other is to embed the credit setting in the credit spread enabled discounting. We will concentrate on the second approach for the time being.

First, we assume that all credit discounting spread from market implied CDS spreads. Given default probability $P_d$ from market CDS spreads, the discounting spread is given by:

$$e^{-(r+sp)t} = e^{-rt}\left[1-(1-R)P_d\right] \tag{3}$$

where the sp is the additional spread needs to be added to the riskfree rate for discounting, and $P_d$ is the cumulative default probability. If only default is concerned, the step by step discounting mechanism is quite simple: if the netting value of the portfolio is positive, which means the counterparty owes money to the concerned financial institution, counterparty's credit spread will be used for additional discounting in that particular scenario and time step; on the other hand, if the netting value is negative, which means the financial institution owes money to the counterparty, the institution's own credit spread should be used for the additional discounting.

Similar concept can be extended to account for credit migration. Assume a reasonable transition matrix is available[3]. By propagating the transition matrix through time, we will have the probability of any trading party at a specific rating at a given time. In another words, the probability of a particular trading party A at $t_i$, scenario j and rating k: $P_{ijk}$ [4].

This probability will allow us to be able to discount cashflows under different ratings:

$$CF_{i-1,j} = \sum_{k} P_{ijk}(Max[CF_{i,j}, Thr_k]D_k + (CF_{i,j} - Max[CF_{i,j}, Thr_k])D_0) \quad (4)$$

where $D_k$ is the discounting corresponding to rating k, $CF_{i,j}$ is the cashflow(or MTM) at time $t_i$ and scenario j, and *Thr* is the rating based threshold. Here we account for the rating based threshold by applying two discount factors, risk free and risky, to the corresponding exposures separately. This of course implies that the discounting will have to be applied after all exposures with a counterparty has been netted. On the tree, the simplest approach would be to assume that there is no correlation between credit migration and the market risk factors. Therefore, the rating distribution at a tree node only depends on the time. $P_{ijk}$ is reduced to $P_{ik}$, and at any given time point can be obtained by propagating the transition matrix.

There are now two credit rating distributions $P_{ijk}^{cpty}$ and $P_{ijk}^{self}$ at every time point, one for the counterparty and one for the considered financial institution. At any given point in time during the backward induction pricing, the MTM values of all financial instruments in the portfolio will be netted first, and a decision will be made in terms of the spreads to be used for credit discounting based on the sign of netted MTM. If MTM is positive, $P_{ijk}^{cpty}$ will be used for discounting, otherwise $P_{ijk}^{self}$ will be used. In later sections, we will discuss credit simulations that would achieve the above objective.

Now we are ready to price all instruments through the pricing framework, whether it is HW trees/lattice or LMM scenarios. There are generally four logic steps at any given time:

- calculate MTM for all financial instruments

- make exercise/trigger decision and calculate option value

- net all MTM values and calculate exposure by applying contractual agreements

---

[3] There are in fact many practical issues in terms of obtaining an accurate credit transition matrix, the discussion of which will deferred to later sections.

[4] Of course, this is based on the assumption that transition matrix is time invariant.

- Applying discounting backward to next time step

At maturity, all the cashflows for all instruments are known with certainty with certainty and there is no exercise decision to make. Once all the cashflows are netted, it will be obvious which spread need to be applied for discounting the cashflows: if the MTM of the portfolio is positive, counterparty's discount spread with different credit ratings will be applied; if the MTM of the portfolio is negative, the concerned financial institution's discount spread with different credit ratings will be applied. At each time step, exercise/trigger decision will also be made for individual trades. Following that, netting will be performed for the portfolio of instruments.

Following the above steps backward to time zero, the pricing of a portfolio of instruments in the presence of default risk with important features in CSA accounted for is obtained.

**4 Simulating Credit**

We now discuss the first approach: credit simulation. Credit simulation not only gives us a way to examine default and ratings migrations directly in scenarios, it also provides us a tool to reflect the correlation between market risk factors and credit defaults/migrations. Given the simulated credit in scenarios, one would be able to account for all credit related events. Again, it is assumed a reasonable transition matrix has been obtained, which provides the right default and migration information over time.

One way to perform credit simulation is based on the structural model, which was first proposed by Merton [MER74]. In a structural model, default occurs when the underlying asset value decreases below than the liability threshold, similar to corporate default behavior.

The first step of credit simulation process would be calibration of the default threshold to CDS market. For each individual counterparty, we need to determine the threshold for default at each point in time. Similarly, rating migration can be achieved by introducing rating thresholds for the underlying asset value. With the probability of transiting to different rating levels readily available from transition matrix, these rating thresholds can also be determined easily[5].

For calibration, one can generate a large number of scenarios through each time steps following simple Geometric Brownian Motion (GBM),

$$dA = \sigma A dw \qquad (5)$$

Where A stands for asset value and $\sigma$ is the volatility of the asset returns. At a specific time, the probability of the reference entity at a particular rating level (including default) corresponds to a range of returns. By examining the returns across scenarios, one can find the relevant rating thresholds. The thresholds are determined so that if the asset return crosses these particular levels, it should result in result in a corresponding rating change. These thresholds will be calibrated for each credit entity j and at all reference times $t_i$, denoted as $H_{ij}$.

---

[5] Technically, the approach we are taking is not a true structural model, as the asset values were never really determined. Rather, we are following the conceptual framework using a dummy variable representing asset value. Different thresholds on asset returns for defaulting and transition are determined based observed market data.

Correlations between the credit and the market risk factors can be achieved through the following structural model:

$$d\mathbf{M} = \mathbf{c}^s \cdot \mathbf{dw}^s + \mathbf{c}^m \cdot \mathbf{dw}^m \qquad (6a)$$

$$d\mathbf{A} = \mathbf{b}^s \cdot \mathbf{dw}^s + \mathbf{c}^A \cdot \mathbf{dw}^A \qquad (6b)$$

$$\rho(M_i, A_j) = c_i^s b_j^s, \quad \rho(A_i, A_j) = b_i^s b_j^s, \quad \rho(M_i, M_j) = c_i^s c_j^s \qquad (6c)$$

Here **A** again is the asset values and **M** is the market risk factors that we want to have correlation with credit risk factors, c and b matrix are the coefficients or factor loading on the random factors. $\mathbf{dw}^s$, $\mathbf{dw}^m$, and $\mathbf{dw}^A$ are for systematic, idiosyncratic market risk and idiosyncratic credit risk factors. $\rho$ are the correlations among market and credit risk factors. Given the already calibrated single name thresholds, the credit states for all scenarios can be known easily.

The above formulations are not limited to the normal distribution framework. One can extend the normalized random numbers to fat tailed distributions, which would allow better modeling of jump events. The single name calibration under normal distributions can be translated into fat tailed distributions by conversion of cumulative distributions:

$$H_{FT} = F^{-1}(N(H_N)) \qquad (7)$$

Where N() is the cumulative normal distribution function, $H_N$ is the threshold under normal distribution and F() is the cumulative fat tail distribution, so $F^{-1}()$ is the inversion of it.

The above formulation will also allow changes of CDS spreads during simulations, indicating the change of credit environment. The easiest way is to make the default thresholds stochastic, for example a mean reverted process:

$$dH = k(a - H)dt + \sigma dw \qquad (8)$$

where a is the mean threshold and k is the mean reversion speed. With H being stochastic, the asset will move closer or further from the default boundary, showing worse or better credit situations.

As was discussed earlier, credit exposure from CVA can be hedged by trading CDS dynamically. To properly reflect the economic values of CDS hedge and CVA, we would need to consistently account for the default risk in both CDS valuation and CVA calculation. This can be achieved with a consistent credit simulations framework.

One complication is the CVA calculation for CDS contract. We know that the CVA for CDS in a simulation framework comes from the joint default of counterparty and reference entity. Therefore we need to simulate joint defaults in order to have accurate CDS CVA calculations. The same is true for CDOs and related credit products, where joint defaults are important.

One key input for simulating joint defaults is the default correlation. However, there is no product traded in the market directly linked to joint default, therefore there is no real market data to calibrate the joint default distribution. One likely possibility is to calibrate to market traded index (CDX, ITRAXX) tranches using structural models (not reduced form copula). This topic deserves will be left for future investigation..

In addition to simulate credit events in Monte Carlo scenarios, it is also possible to generate credit states within pricing trees/lattices. With credit states and probabilities generated, CVAs can be calculated in exactly the same way within trees/lattices as in Monte Carlo simulations. We will use tree as an example.

For a tree, we have the probabilities for each node $j$ at every point in time $i$: $P_{ij}$, and we have $\sum_j P_{ij} = 1$. If we assume there is no correlation between market risk factors and the credit default, it would be trivial to determine the joint credit and market moves, as it is simply the cross product of credit migration and the market risk trees. It is however a littlie bit trickier when there is correlation.

With correlations between market risk factors and credit events known, the credit information at every node can be generated using a three-step process. In the first step, we will look at the distribution of market risk factors, and the calibrated credit risk factors, so we have all the necessary distribution information, such as return percentiles. For example, at a given time $t_i$, we have

$$S_0, S_1, S_2, ..., S_N$$

for spot values, and the corresponding probability at each node:

$$P_0, P_1, P_2, ..., P_N$$

That gives a discrete distribution P(S) for the market risk factors. This would be the same distribution if we have used Monte Carlo simulation and have followed the same risk neutral process. In the second step, we would generate the correlated random numbers for all time steps given the correlations between market risk factors and credit risk factors. This can be achieved either with factor models or taking the square root of the correlation matrix[6]. The third step would involve the mapping of correlated random numbers into individual credit states: The market risk random numbers will be mapped to the tree nodes given the percentile distributions, and the credit states can be generated for each node by looking at the corresponding credit risk random numbers. Given the credit states for each node, one can evaluate the exposure for relevant credit states.

---

[6] Either Cholesky decomposition or singular value decomposition can be used for this purpose.

Another possible approach to generate credit information using tree is to create correlated tree with both assets and market risk factors, which likely will be a quite involved process. With all asset information known for every tree node, the credit states can be calculated from calibrated threshold information.

With simulated credit scenarios, forward simulation framework can be adjusted to manage rating based collateral thresholds and ATEs. MTM will only need to be calculated for defaulted scenarios, with exposures calculated based on rating based threshold.

**5 Risk Neutralized Transition Matrix**

Historical transition matrix, propagating in time, can provide time dependent rating migrations information. However, the defaults generated from the propagation in general do not match the risk neutral default probabilities derived from CDS market. The reason for this discrepancy is obvious: the historical default behavior is different from what's predicted in the market place for future, and does not reflect current credit environment. In addition, market implied default rates also include some risk premium in it.

To consistently price credit risk, we need risk neutralized transition matrix for our credit simulations, just as market traded instruments are priced in a risk neutral measure. McNulty and Levin [NV00] proposed a way to translate historical transition matrix to risk neutral transition matrix based on CAPM model. In their methodology, they assumed that asset returns in real/historical world can be translated into risk neutral world by introducing a risk premium term $\rho\theta$, where $\rho$ is the correlation between asset and market. Assuming the asset returns are normally distributed, the risk neutral probability can be therefore derived with a simple risk premium term:

$$P_{RN}(R < b) = N(b + \rho\theta) \quad (9)$$

Where N() is the cumulative normal distribution function. Using this translation, they were able to link the historical transition matrix to the market traded defaults, which is implied in CDS trading.

McNulty's approach is simple and gives approximate defaults in the risk neutral world, however, may not be consistent with all default information from market. For our CVA calculations, we need a more accurate transition matrix, which is consistent with available CDS term structures, so that the scenario defaults are reliable.

In developing the term structure model of credit spreads, Jarrow, Lando and Turnbull [JLT97] used a transition matrix approach with matrix parameters calibrated to the market traded spreads. The calibration is done by applying risk premium to each rating and time periods so to match the market. This process is done year over year with different risk premium at different time, implying the transition matrix will not being Markovian. The following shows the transition matrix from t to t+1:

$$\tilde{Q}_{t,t+1} = \begin{pmatrix} \tilde{q}_{11}(t,t+1) & \tilde{q}_{12}(t,t+1) & \ldots & \tilde{q}_{1n}(t,t+1) \\ \tilde{q}_{21}(t,t+1) & \tilde{q}_{22}(t,t+1) & \ldots & \tilde{q}_{2n}(t,t+1) \\ \ldots & \ldots & \ldots & \ldots \\ 0 & 0 & \ldots & 1 \end{pmatrix}$$

where tild indicates in the risk neutral world. The linking between historical world and risk neutral world is through:

$$\tilde{q}_{ij}(t,t+1) = \pi_i(t) q_{ij}(t,t+1)$$

with $\pi_i(t)$ being the risk premium applied to the historical transitions. Thus JLT's idea is to start with a historical transition matrix and apply risk premium as necessary to make the matrix risk neutral over time.

Even though the historical transition matrix is arguably difficult to obtain in the first place for different trading entities, such as corporates vs. financials vs. insurance companies etc, we feel it is prudent to make no assumption, such as non-Markovian property in JLT, about what we start with and derive a transition matrix. It is also interesting to see to what extent market traded CDS spreads can be accounted for with a Markovian transition matrix. To this end, we decide to derive the transition matrix by optimizing (constrained) the matrix variables, so that the propagated default probabilities are consistent with the default probabilities implied from market.

Below is a simple example to illustrate the derivation of the risk neutralized transition matrix (RNTM). Assume you have a transition matrix of 4x4 with A,B,C,D rating and the market traded default probability (based on recovery assumption) at different point of time is

| P(Default) | A | B | C |
|---|---|---|---|
| 1 | 0.31% | 1.72% | 6.28% |
| 2 | 0.72% | 4.27% | 11.80% |
| 5 | **2.60%** | **13.00%** | **25.60%** |
| 10 | 7.00% | 30.00% | 48.00% |

Table 1

Where the default probabilities for ratings A, B, C at listed for terms 1, 2, 5, and 10 yrs. Then the 3x3 matrix excluding the default row and column can be used as parameters in fitting to the default matrix. The optimized transition matrix

| Transition Matrix | A | B | C | D |
|---|---|---|---|---|
| A | 96.00% | 2.50% | 1.19% | 0.31% |
| B | 0.40% | 83.00% | 14.87% | 1.73% |
| C | 0.41% | 1.00% | 92.30% | 6.29% |
| D | 0.00% | 0.00% | 0.00% | 100.00% |

should recover all the market traded default probabilities within reasonable range, and therefore are deemed to be risk neutral. In this case, we say the transition process is almost perfectly Markovian.

However, iteratively propagating transition matrix forward assumes that every year credit migration and default would stay the same. The default/transition intensity is therefore not going to change for a specific rating. On the other hand, market implied default probability reflect different views about default intensities for different time periods (terms), which could cause problem in fitting the CDS implied defaults probabilities with the risk neutralized transition matrix. For a more realistic transition matrix with many ratings, the optimization of risk neutralized transition matrix is a even more complicated problem. In practical implementations, we use Levenberg-Maquart algorithm to optimize the transition matrix parameters in a multi-dimensional space.

In general we can consider the market traded defaults as:

Market Implied Defaults = Transition Matrix Migrations + Perturbations from Average Long Term View

The perturbations from long term view includes the specific risk of the particular company from the average rating view and noises in the market placec due to liquidity. They could be real, or simply noise, which may be used in making trading decisions. In the following we give an example of our risk neutralization process for a more realistic matrix and the deviations of migration vs market traded CDS.

The risk neutralization of transition matrix is done through an optimization process in a multi-dimensional space. We achieve the optimization through the Levenberg-Marquardt algorithm. As an example, we have the probabilities of default term structure estimated as the following:

| P(Default) | AAA | AA | A | BAA | BA | B | C |
|---|---|---|---|---|---|---|---|
| 1 | 0.51% | 0.81% | 0.92% | 1.25% | 2.87% | 6.67% | 12.50% |
| 2 | 1.21% | 1.93% | 2.24% | 3.06% | 7.12% | 13.85% | 24.52% |
| 3 | 2.13% | 3.36% | 3.94% | 5.38% | 12.36% | 24.19% | 37.30% |
| 5 | **4.33%** | **6.75%** | **7.89%** | **10.74%** | **21.94%** | **41.23%** | **56.80%** |
| 7 | 6.37% | 9.97% | 11.69% | 15.80% | 31.00% | 55.46% | 70.68% |
| 10 | **9.57%** | **14.96%** | **17.83%** | **23.79%** | **44.54%** | **72.65%** | **87.80%** |
| 15 | 14.37% | 23.74% | 27.89% | 36.21% | 58.11% | 83.77% | 97.28% |
| 20 | 20.11% | 32.38% | 37.46% | 48.07% | 72.34% | 97.30% | 99.00% |
| 25 | 25.05% | 39.91% | 46.12% | 58.85% | 84.17% | 99.00% | 99.00% |
| 30 | 28.24% | 46.17% | 53.70% | 68.20% | 93.18% | 99.00% | 99.00% |

Table 3

In our optimization, we also use a weighted scheme, where the weightings of short/long term defaults are different: there is virtually no long term default trading in the market, therefore we put in much less weights on the long term points. The optimization parameters are the transition matrix parameters from AAA to C, which totals 49 parameters. The following gives one optimized transition matrix, which does a good job in replicating most of the default information:

| Year 1 | AAA | AA | A | BAA | BA | B | C | D |
|---|---|---|---|---|---|---|---|---|
| AAA | 89.13% | 9.76% | 0.46% | 0.08% | 0.03% | 0.07% | 0.10% | 0.37% |
| AA | 0.97% | 87.58% | 9.61% | 0.51% | 0.33% | 0.18% | 0.21% | 0.61% |
| A | 0.22% | 2.35% | 86.92% | 6.98% | 1.54% | 0.78% | 0.51% | 0.69% |
| BAA | 0.37% | 0.66% | 4.00% | 81.64% | 8.20% | 3.30% | 0.78% | 1.05% |
| BA | 0.10% | 0.30% | 0.50% | 2.00% | 80.97% | 11.00% | 3.07% | 2.05% |
| B | 0.06% | 0.07% | 0.07% | 0.07% | 0.50% | 75.00% | 19.70% | 4.53% |
| C | 0.10% | 0.10% | 0.10% | 0.10% | 0.10% | 0.10% | 93.00% | 6.40% |
| D | 0.00% | 0.00% | 0.00% | 0.00% | 0.00% | 0.00% | 0.00% | 100.00% |

Table 4

This matrix gives the following probability of default:

| Fitted P(D) | AAA | AA | A | BAA | BA | B | C |
|---|---|---|---|---|---|---|---|
| 1 | 0.65% | 1.07% | 1.20% | 1.51% | 3.51% | 7.56% | 14.65% |
| 2 | 1.37% | 2.20% | 2.51% | 3.30% | 7.46% | 15.96% | 27.15% |
| 3 | 2.16% | 3.39% | 3.92% | 5.35% | 11.80% | 24.54% | 37.82% |
| 5 | **3.92%** | **5.96%** | **7.12%** | **10.21%** | **21.23%** | **40.70%** | **54.67%** |
| 7 | 5.93% | 8.80% | 10.79% | 15.81% | 30.91% | 54.44% | 66.94% |
| 10 | **9.38%** | **13.55%** | **17.04%** | **24.90%** | **44.41%** | **70.05%** | **79.37%** |
| 15 | 16.20% | 22.55% | 28.49% | 39.62% | 61.94% | 85.52% | 90.55% |
| 20 | 24.08% | 32.24% | 39.81% | 52.00% | 73.39% | 93.00% | 95.62% |
| 25 | 32.55% | 41.82% | 49.98% | 61.67% | 80.60% | 96.53% | 97.93% |
| 30 | 41.09% | 50.71% | 58.65% | 69.09% | 85.24% | 98.19% | 98.98% |

Table 5

There are two aspects when we link this matrix and the specific names. First in practice, the specific name CDS will be different from the average rating default. Two companies with the same rating could have quite different traded CDS spreads. The second is the deviations of fitted transition matrix default from the market traded defaults, which could be regarded as noise from market liquidity as well as assumption breakdown of transition matrix model: The default and transitional intensity may not be regarded as constant through time, and there are residue non-Markovian characteristics in the real transition process.

In implementing the risk neutralized transition matrix in the CVA calculations for specific names, the generated ratings distribution can be adjusted to accommodate the market traded default probabilities. At a specific point in time, the default probabilities from transition matrix propagation will be set equal to the market traded default, whereas the rest of the ratings populations will be scaled accordingly. This can be implemented in the single name calibration process, where the default threshold will be adjusted to match the market traded defaults. This adjustment would ensure the individual defaults are matched exactly and CDS prices will be recoverd in the valuations.

Risk neutralized transition matrix with ratings migration could also provide some insight into the underlying dynamics implied by CDS market. While most of defaults are through transitions,

RNTM provides a nice way to examine the economics implied from the CDS spreads. One can decompose the CDS spread several components. From transition matrix computations, one can derive, for example out of AA 10yr total default 15%, how much it is coming from transitional default through a specific rating and how much is coming from jump to default. This line of thinking could provide a useful view and another layer of hedging strategy to hedge generic market credit moves.

## 6 A More Efficient Framework

In this section, we will present a new CVA calculation methodology based on backward pricing framework, but from a different perspective:

$$CVA = E^Q \left[ (1-R) D_t \min(Threshold, V_t)_{|A<H_d} \right] \tag{10}$$

where A is asset value, H is the default threshold, R is the recovery assumption, D is the discounting and Q indicates it is risk neutral measure. In this definition, CVA is the integration of all discounted risk neutral default exposures from current time to maturity. What is different from the backward pricing in previous sections is that implied in the above formula we are no longer valuating the portfolio twice anymore, once is sufficient: we are only concerned about the defaulted scenarios, and CVA is calculated by collecting the defaulted values through time. We will use Monte Carlo simulation as an example. Tree/lattice implementation would be trickier if rating migration is considered.

In general the process would involve three different steps:

    Scenario generation (market and credit)

    Valuations under the market scenarios, and save all values

    Collect CVAs by aggregating defaulted exposures (market by credit)

The first step has been described in details in previous sections.

In the second step, deal is valued exactly the same way as what one would usually do in regular valuations, except that it is done with the same valuation framework: generic time grid, generic term structure and market rates, which are all embedded in the market scenarios. As in the backward pricing framework, valuations of all instruments are done with the same pricing trees/lattices/MC scenarios. The valuations of each deal can be separate or aggregated. Once the valuations are done, their values at the generic time grid and each scenario will be saved in a database. For credit instruments like CDS, one would also need the credit scenarios in valuations.

When the valuations are done separately for each deal and each market scenario set, the computations can be distributed on a computer grid easily as the pricing of every deal per scenario set is a separate computer job. While the memory and computation requirement could be intensive for large portfolios, we can resort to some simple tricks. One way to save time and memory is to work with cashflow instruments more efficiently. For simple linear instruments with only cashflows, all cashflows could be aggregated together as one cashflow instrument with many cashflows along time. For example, if there 1000 swaps with one counterparty, all these swaps can be collapsed into one single cashflow instrument with cashflows at many time points. This would reduce the requirement for memory and computations: only one value, instead of 1000 values, for each scenario and time step needs to be stored in memory. There certainly could be some approximations required for distributing remote cashflows. For non-linear instruments, it is also possible to collapse multiple instruments into one generic instrument. For example, the caps can be collapsed into one cap as caplets are separable. One can also convert daily digital caps into weekly or even monthly ones, which would save computations significantly.

The third step is the aggregation of CVAs. This step is the overlap of credit and market scenarios. Based on the credit scenarios we generated in the previous section, we would know the credit state of a trading party at time $t_i$ and scenario j. In each of the scenarios, we would first apply the trigger events: if termination event is triggered without default, the aggregation would discontinue. Since we are only collecting defaulted exposures in this framework, we would look for all defaulted scenarios for both trading parties, in which we net all the trade values and apply rating based thresholds to calculate the defaulted exposure. The exposure at a given time would be:

$$\max(0, \min(H, NetValue * defaultFlag)) \qquad (11)$$

where the default flag is 1 if counterparty defaults and -1 otewise. So when counterparty defaults and netted value is negative, there is no credit exposure. In applying the threshold, we also need to go backward one step to find the rating of the trading party right before default, which is used in looking up the rating based threshold table.

The most significant advantage of this new framework is the efficiency and computational flexibility. Every deal valuation for a specific scenario set is one separate computer job, which can be distributed on a computational grid. For each CVA calculation, every deal has to be valued only once within the generic pricing model setup. The calculations of incremental CVA, market/credit risk greeks and evaluations of wrong way exposure can be achieved with great efficiency.

Another important advantage of this new framework over the regular backward pricing framework is the accuracy gained when calculating CVA for exotic structures, in which significant amount of regressions have to be performed in determining the exercise boundary if Longstaff –Schwartz algorithm is used [LS98]. While the regressions under risky and riskless settings can be significantly different, the noise from the regression in practice can be quite significant as we are taking the difference of two big numbers. This is especially the case when we are limited to small number of simulation paths for the reason of practical cost. With the new framework, the regression noise issue is avoided by aggregation, where only defaulted exposures are collected.

To achieve better convergence, for example in dealing with joint defaults, one may need a lot of credit scenarios. This does not cause problem in the current framework. In the third aggregation step, one can aggregate for example 100,000 credit scenarios with 10,000 saved market scenario values. So every market scenario will have 10 credit scenarios correspond to it, as if every market scenario has a credit distribution.

Assuming there is no CVA for CSA with zero collateral threshold, the ATE can also be treated as being zero-threshold.

**7 Incremental CVA, Greeks Computations and Wrong Way Exposure Estimation**

It is straightforward to calculate incremental CVA efficiently, given the saved portfolio values at every market scenario and the generated market/credit scenarios. With a new trade coming in, the trade will be priced within the same market scenarios, and the result will be aggregated along with the saved scenario values. The difference in new total CVA and old CVA would be the incremental CVA.

For the hedging of CVA, greeks are needed in terms of market risk factor sensitivities like delta, gamma, vega etc, as well as the credit risk sensitivities like the CDS delta. With the flexible distribution of computation tasks, the computation of market risk sensitivities are simply new market scenario generation, whereas the credit risk sensitivities means new credit scenarios based on changes in specific CDS spread. In calculating the CDS sensitivity, the CDS spread shift can be achieved by re-calibrating the default thresholds for the specific name and re-generate the credit scenario set. With all valuations per deal and per market scenario set calculated separately on the distributed computational grid, sensitivity calculations can be achieved with great efficiency.

As credit scenarios are generated with different correlation assumption between the credit and the market risk factors, wrong way exposure can also be evaluated by changing the correlation assumption, which would give different CVAs showing the likely co-jump in default and market risk factors. This can be accomplished with the same market scenarios: the random numbers that are used in generating the credit scenarios can be re-mapped to reflect the new correlation. This would, by design, save additional computations from market scenario changes.

**8 Exercise Boundary of Options in the Presence of CVA**

The above approach makes assumption that the exercise decision in the valuation of options will not be affected by the credit exposure or CVAs. This is mostly due to the fact that we want to be able to compute the CVA more efficiently through distributed computing, where deal by deal valuations and credit aggregations are done at different stages, and are different jobs in implementations. As such, the option exercise boundary is not regressed based on future potential defaults. So how significant is this approximation?

For a typical Bermudan swaption, which physically settles, meaning the exercise would lead to an underlying swap. The decision of whether exercise the option is based the comparison of two quantities: the value of future options and the current value of underlying swap. If current value of underlying swap (meaning exercising) is greater than the future option value (meaning not exercising), then you would exercise the option, and vice versa. The counterparty credit implication on this exercise decision lies in the difference in credit value adjustment for underlying swap and future option.

Now let us explore it with a specific example: a 20y Bermudan 4.5% fixed coupon vs. libor flat right before an exercise date. Based on a low interest rate environment (we used 2010/09/10 market data for this exercise) without any credit considerations, the optimal decision is to exercise the option. The following table shows the change of exercise boundary when counterparty credit risk is considered. We compared three different possible default risk levels (0%, 1%, and 2% annual default rate). The exercise boundary changes by roughly 20bps for counterparty with 1% annual default, and dropped another 20bps if the counterparty has 2% annual default rate.

| Scenario | Exercise Boundary |
|---|---|
| No Credit | 4.5% |
| 1% annual Default | 4.3% |
| 2% annual Default | 4.1% |

Table 6

The reason for the change of exercise boundary is quite obvious: the option value is one sided with values localized at low libor rates and also tends to be longer term than the underlying swap, where the fix/float coupons cashflow exchanges are two sided and are more front loaded. Therefore we would expect the counterparty default effect is less for the underlying swaps than the future options. Hence more CVA effect from the future option value, meaning future option value drops more than the underlying swap when default is considered. And the exercise boundary drops as a result. Intuitively, the exercise boundary behavior would be different if the curve shape changes from steep upward sloping to flat and to downward sloping shape. For a downward sloping curve, the option values will be even more backend loaded and the difference between the future option and underlying swap CVAs could be even greater.

The next question naturally is what kind effect this change of exercise boundary would have on the CVA calculation of this Bermudan swaption structure.

For a 10M 20yr no call 6 month Bermudan swaption with 5% fixed coupon vs libor flat, the CVA is around $305K for a default probability of 2% per year. The change of exercise boundary effect due to the presence of CVA is around $1800. For a no call 2y Bermudan swaption, the CVA exercise boundary effect is $6000 out of $330K CVA. So this is rather a small effect. This is expected since only the states around the exercise boundary would be impacted, which generally have rather small or close to zero values. The rest of the exercise decisions for a majority of the distribution would remain the same.

The exercise boundary would also change when the netting of portfolio is considered and when additional credit provisions of CSA are included in the calculation. For example, considering one particular option by itself may indicate that economically it should be exercised; including the netting with other positions with the same counter party might indicate the overall position is more an asset rather than a liability, which would then change the exercise boundary and exercise decision.

The correlation between market risk factor and asset default would obviously have significant effect on the exercise boundary as well. It is expected the exercise boundary regression will be skewed by defaults, meaning the option exercise decision would be affected by the changes of asset values (a proxy for credit state) as well. Therefore, If Longstaff Schwartz regression is utilized to determine the exercise boundary, one would include the structural asset variables among the explanatory variables. In such cases, the CVA would be affected by the above assumption.

Within the current framework, we are ignoring the effect of CVA on exercise boundary, which are small most of the time. However, it is always prudent to understand the characteristics of the products, portfolios and market environment, as well as the effects of such assumptions. This can be achieved by comparing results from rigorous backward pricing framework with and without credit included decision making. In the case of sizable impact, one may take additional corrections in exercising the option. We will leave this for future explorations.

## 9 Practical Issues

There are some practical issues that are relevant to the realization of all economic values from the CVA calculations. For example, in order to realize the values from automatic trigger event, one would need to know as soon as the downgrade event happens, and then make decision about exercising the trigger option. The same is true for mutual put breaks. Mutual puts are the same as a series of Bermudan options, which should be executed as if they are the same as normal options with exercise decisions. Missing the exercise decision of such option when counterparty owes a lot of money could mean significant loss economically.

In practice ATE and Mutual put breaks may not get executed economically because of practical reasons. For example, relationship between the two trading party, or the economic value is small and there is not enough incentive to call off the deal. However, the economic value of these options is clearly there and the execution issue is more of practical decision, not economic issue. In CVA calculations, one could put in a barrier, below which ATEs and MPs will not be executed. Thus simulate the real situation.

## 10 Conclusion

We have reviewed several approaches for CVA calculation, and compared the advantages and disadvantages of these methods. We proposed an efficient computation framework, which can be easily implemented for distribution computing. We also touch on potential impact of counterparty risk on option exercising decision.